\documentclass[12pt]{article}
\usepackage{amsmath,amssymb,url,epsfig,color,cite}

\definecolor{verdes}{cmyk}{0.92,0,0.59,0.4}
\ifx\pdfoutput\undefined
\usepackage[dvips,bookmarks=false]{hyperref}	
\else
\usepackage{hyperref}	
\fi
\hypersetup{colorlinks,bookmarksopen,bookmarksnumbered,citecolor=verdes,
linkcolor=blus,pdfstartview=FitH,urlcolor=rossos}

\def\m@th{\mathsurround=0pt }
\def\leftrightarrowfill{$\m@th \mathord\leftarrow \mkern-6mu
        \cleaders\hbox{$\mkern-2mu \mathord- \mkern-2mu$}\hfill
        \mkern-6mu \mathord\rightarrow$}

\def\overleftrightarrow#1{\vbox{\ialign{##\crcr
        \leftrightarrowfill\crcr\noalign{\kern-1pt\nointerlineskip}
        $\hfil\displaystyle{#1}\hfil$\crcr}}}

\definecolor{rosso}{cmyk}{0,1,1,0.4}
\definecolor{rossos}{cmyk}{0,1,1,0.55}
\definecolor{rossoc}{cmyk}{0,1,1,0.2}
\definecolor{blu}{cmyk}{1,1,0,0.3}
\definecolor{blus}{cmyk}{1,1,0,0.6}
\definecolor{bluc}{cmyk}{1,1,0,0.1}
\definecolor{verde}{cmyk}{0.92,0,0.59,0.25}
\definecolor{verdec}{cmyk}{0.92,0,0.59,0.15}
\definecolor{verdes}{cmyk}{0.92,0,0.59,0.4}
\definecolor{grigio}{cmyk}{0,0,0,0.07}
\definecolor{rosa}{cmyk}{0,0.1,0.1,0.02}
\definecolor{rosino}{cmyk}{0,0.05,0.05,0.02}
\definecolor{rosas}{cmyk}{0,0.3,0.25,0.05}
\definecolor{celeste}{cmyk}{0.1,0,0,0.02}
\definecolor{giallino}{cmyk}{0,0,0.4,0.02}
\definecolor{rosso}{cmyk}{0,1,1,0.4}
\definecolor{rossos}{cmyk}{0,1,1,0.55}
\definecolor{rossoc}{cmyk}{0,1,1,0.2}
\definecolor{blu}{cmyk}{1,1,0,0.3}
\definecolor{bluc}{cmyk}{1,1,0,0.1}
\definecolor{blucc}{cmyk}{0.7,0.5,0,0}
\definecolor{viola}{cmyk}{0,1,0,0.6}
\definecolor{viola2}{cmyk}{0,1,0.2,0.6}
\definecolor{verde}{cmyk}{0.92,0,0.59,0.25}
\definecolor{verdec}{cmyk}{0.92,0,0.59,0.15}
\definecolor{verdes}{cmyk}{0.92,0,0.59,0.4}
\definecolor{verdino}{cmyk}{0.12,0,0.09,0.05}
\definecolor{giallo}{cmyk}{0,0,1,0}
\definecolor{gialloverde}{cmyk}{0.44,0,0.74,0}

\def\simlt{\stackrel{<}{{}_\sim}}
\def\simgt{\stackrel{>}{{}_\sim}}

\font\tenrsfs=rsfs10 at 12pt
\font\sevenrsfs=rsfs7
\font\fiversfs=rsfs5
\newfam\rsfsfam
\textfont\rsfsfam=\tenrsfs
\scriptfont\rsfsfam=\sevenrsfs
\scriptscriptfont\rsfsfam=\fiversfs
\def\mathscr#1{{\fam\rsfsfam\relax#1}}

\newcommand{\be}{\begin{equation}}
\newcommand{\ee}{\end{equation}}
\newcommand{\beq}{\begin{equation}}
\newcommand{\eeq}{\end{equation}}

\def\shat{\ifmmode \hat{s}\else $\hat{s}$\fi}
\def\gp2{{g'}^2}
\def\g2{g^2}
\def\g32{g_s^2}

\newcommand{\newc}{\newcommand}

\newc{\gsim}{\lower.7ex\hbox{$\;\stackrel{\textstyle>}{\sim}\;$}}
\newc{\lsim}{\lower.7ex\hbox{$\;\stackrel{\textstyle<}{\sim}\;$}}
\newc{\ie}{{\it i.e.}}
\newc{\etal}{{\it et al.}}
\newc{\mev}{\hbox{\rm\,MeV}}
\newc{\tev}{\hbox{\rm\,TeV}}
\newc{\xpb}{\hbox{\rm\, pb}}
\newc{\xfb}{\hbox{\rm\, fb}}

\newc{\G}{{\cal G}}
\newc{\h}{{\cal H}}
\newc{\D}{{\cal D}}
\newc{\E}{{\cal E}}

\newc{\mtop}{M_t}
\newc{\mbot}{m_b}
\newc{\mz}{M_Z}
\newc{\mw}{M_W}
\newc{\alphasmz}{\alpha_s(M_Z)}
\newc{\swsq}{\sin^2\theta_W}
\newc{\cwsq}{\cos^2\theta_W}
\newc{\tw}{\tan\theta_W}
\newc{\cw}{\cos\theta_W}
\newc{\sw}{\sin\theta_W}
\newc{\BR}{\hbox{\rm BR}}
\newc{\zbb}{Z\to b\bar}
\newc{\Gb}{\Gamma (Z\to b\bar b)}
\newc{\Gh}{\Gamma (Z\to \hbox{\rm hadrons})}
\newc{\sgn}{\mbox{sgn}}

\def\eq#1{eq.~(\ref{#1})}
\def\fig#1{fig.~\ref{#1}}

\def\vev#1{\langle {#1} \rangle}

\newcounter{mysubequation}[equation]

\newcommand{\TeV}{\,\mathrm{TeV}}
\newcommand{\GeV}{\,\mathrm{GeV}}

\def\dm2{\delta m^2}
\def\dv2{\delta v^2}

\def\beq{\begin{equation}}
\def\eeq{\end{equation}}
\def\bea{\begin{eqnarray}}
\def\eea{\end{eqnarray}}
\def\be{\begin{equation}}
\def\ee{\end{equation}}
\def\bea{\begin{eqnarray}}
\def\eea{\end{eqnarray}}
\def\slashchar#1{\setbox0=\hbox{$#1$}           
   \dimen0=\wd0                                 
   \setbox1=\hbox{/} \dimen1=\wd1               
   \ifdim\dimen0>\dimen1                        
      \rlap{\hbox to \dimen0{\hfil/\hfil}}      
      #1                                        
   \else                                        
      \rlap{\hbox to \dimen1{\hfil$#1$\hfil}}   
      /                                         
   \fi}                                         %
\catcode`@=11
\long\def\@caption#1[#2]#3{\par\addcontentsline{\csname
  ext@#1\endcsname}{#1}{\protect\numberline{\csname
  the#1\endcsname}{\ignorespaces #2}}\begingroup
    \small
    \@parboxrestore
    \@makecaption{\csname fnum@#1\endcsname}{\ignorespaces #3}\par
  \endgroup}
\catcode`@=12

\begin{document}

\begin{center}
\hfill
CERN-PH-TH/2012-209

\bigskip\bigskip\bigskip

{\Large\bf\color{black} 
Correlation between the Higgs Decay  Rate \\[3mm] to Two Photons and the Muon $g-2$ }\\

\medskip
\bigskip\color{black}\vspace{0.6cm}
{\bf
\large Gian F. Giudice$^a$, {\bf Paride Paradisi}$^a$ {\rm and} \bf Alessandro Strumia$^{b,c}$}
\\[7mm]
{\it $^a$ CERN Theory Division, CH-1211, Geneva 23, Switzerland}\\
{\it $^b$ Dipartimento di Fisica dell'Universit{\`a} di Pisa and INFN, Italia}\\
{\it  $^c$ National Institute of Chemical Physics and Biophysics, Tallinn, Estonia}\\

\bigskip\bigskip\bigskip\bigskip

{
\centerline{\large\bf Abstract}
\begin{quote}

\medskip
In the context of minimal supersymmetry with slepton mass universality 
we find that an enhancement in $h\to\gamma\gamma$ by at least 40\%, as
hinted by present data, implies a deviation of 
the muon anomalous magnetic moment by exactly the right amount to explain the observed anomaly.
The enhancement in $h\to \gamma \gamma$ selects a light stau with large left-right mixing,
a light Bino, and heavy higgsinos. The corresponding parameters are  compatible with
thermal dark matter, predict small deviations in $h\to Z \gamma$ and $h\to \tau\tau$, and measurable violations of lepton universality.

\end{quote}}

\end{center}

\section{Introduction}

The Higgs boson discovered at the LHC has couplings roughly in agreement with the Standard Model (SM) prediction. At present, deviations from this prediction are too poorly constrained by the experimental data to allow for definite conclusions, but there are indications for an excess of the Higgs rate in the diphoton channel~\cite{exp,fit}. If we take this hint seriously together with the indication that the Higgs rates in $ZZ^*$ and $WW^*$ are consistent with the SM, we are led to the conclusion that the effective coupling between the Higgs and two photons must receive new contributions beyond the SM.

In the context of supersymmetric theories, there are several new particles that affect the Higgs-photon coupling at the quantum level. However, most of them do not lead to the desired effect. Stops give contributions to the Higgs-gluon coupling that overcompensate the effect
in the photon coupling, thus reducing $\sigma(pp\to h){\rm BR}(h\to \gamma \gamma)$. The charged Higgs and charginos give only small effects in the Higgs-photon coupling. Hence, the main supersymmetric candidate for an increased di-photon width is a light stau which, in presence
of a large left-right mixing, increases the Higgs-photon coupling~\cite{hggSUSY}.
An alternative strategy is to invoke supersymmetric contributions to reduce the $hbb$ coupling and consequently 
enhance all other Higgs branching ratios, including $h\to\gamma\gamma$~\cite{newref}. 
By considering cases in which the Higgs pseudoscalar is not too heavy, it is possible to obtain rates for $h\to WW^*,ZZ^*$ similar to those of the SM, together with an enhanced value of $h\to\gamma\gamma$.

In this paper we study the conditions under which a light stau can enhance $h\to \gamma \gamma$, showing that this can happen only for special and extreme values of the supersymmetric parameters. Our most important result is that these special parameters, under the assumption of soft mass universality in the lepton sector, give a strong correlation between a large enhancement of $\Gamma(h\to \gamma \gamma)$ and an increase 
of the anomalous magnetic moment of the muon ($a_\mu$). Whenever $\Gamma(h\to \gamma \gamma)$ is significantly enhanced, the value of $a_\mu$ differs from the SM expectation and, interestingly, turns out to be in  agreement with measurements, explaining the observed discrepancy with the SM~\cite{bnl,gm2mth},
\beq
\delta a_\mu = a_\mu^{\rm exp}-a_\mu^{\rm SM} = (2.8\pm 0.8) \times 10^{-9}.
\eeq
Furthermore, we show that the supersymmetric parameters selected by a large enhancement of $\Gamma(h\to \gamma \gamma)$, beside explaining $\delta a_\mu$, can correctly account for
dark matter with thermal relic abundance, are consistent with electroweak (EW) precision data,
give small effects in $\Gamma (h\to Z\gamma )$ or $\Gamma (h\to \tau\tau )$, and give observable violations of lepton universality.

\section{Enhancing $h\rightarrow \gamma\gamma$}

\begin{figure}[t]
$$
\includegraphics[width=0.5\textwidth]{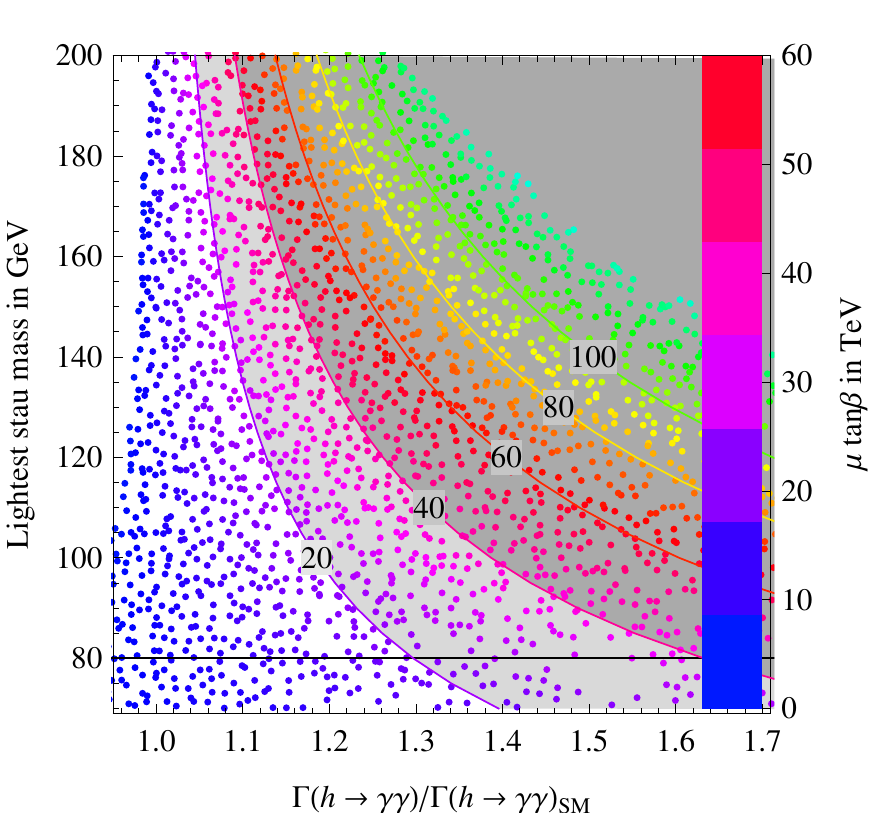}\qquad \includegraphics[width=0.45\textwidth]{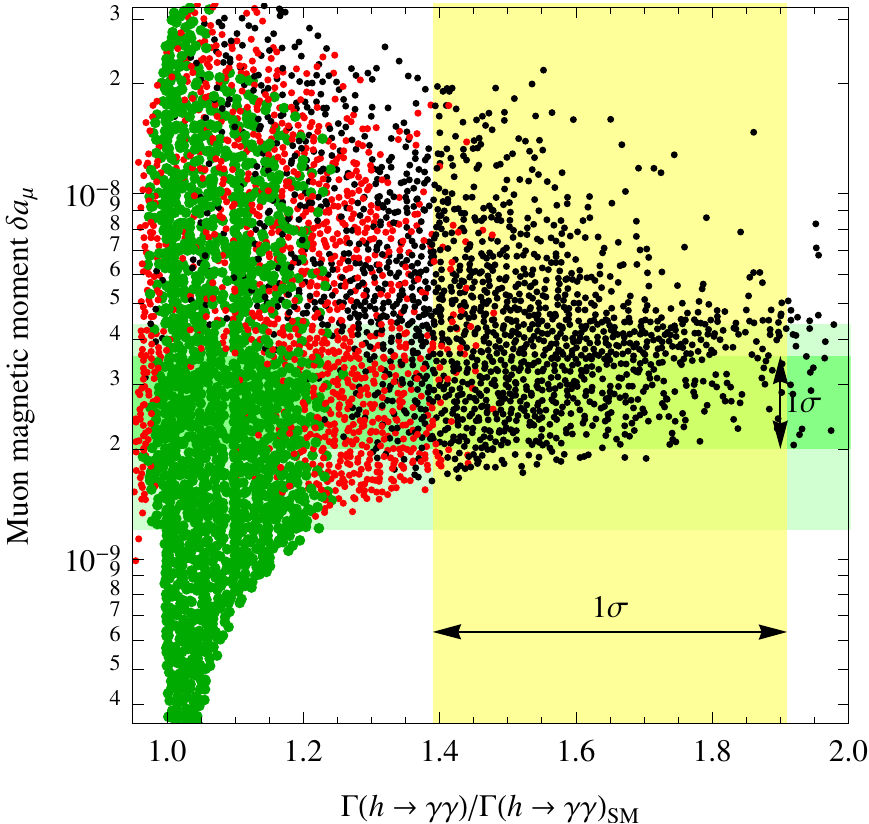}$$ 
\caption{\em {\bf Left:} The value of the lightest stau mass needed to obtain a given
$\Gamma(h\to \gamma\gamma)$ for different values of $\mu\tan\beta$ (denoted by the
color code shown in the figure). The points are obtained through a scan as described
in \eq{scan}.
The contour lines of $\mu\tan\beta$ in {\rm TeV}
are obtained from the approximate expression in~\eq{eq:approx}, which can be trusted
only in the region corresponding to large enhancements of $\Gamma(h\to \gamma\gamma)$.
Vacuum stability bound at tree-level imply $|\mu\tan\beta|  \simlt 40\TeV$. 
{\bf Right}: Correlation between the supersymmetric contributions to the muon $g-2$
and to $\Gamma(h\to\gamma\gamma)$. The bands show the regions favored  by present experimental data. Green (red) dots correspond to a $\tilde\tau$ heavier than $100$ $(80)\GeV$, and black dots correspond to a lighter $\tilde\tau$, which is experimentally allowed only if it is quasi-degenerate to a neutralino. All points satisfy the metastability bound.
\label{fig:gh}}
\end{figure}

The starting point of our analysis is the Higgs decay width into two photons mediated
by $W$, top, and staus~\cite{Djouadi}: 
\be
\Gamma(h \to \gamma\gamma) = \frac{\alpha^{3}m^3_h}{256\pi^2 \sin^2\theta_{\rm W} M^2_W}
         \left| F_{1}\left(\frac{4M_W^2}{m_h^2}\right) +
                N_c Q_t^2 F_{1/2}\left(\frac{4m_t^2}{m_h^2}\right) +
                \sum_{i=1,2} g_{h \tilde{\tau}_i\tilde{\tau}_i}
                \frac{M^2_Z}{m_{\tilde{\tau}_i}^2}
                F_{0}\left(\frac{4 m_{\tilde{\tau}_i}^2}{m_h^2}\right)
         \right|^2\,,
         \label{gamrat}
\ee
where $N_c=3$, $Q_t=2/3$ and the loop functions are
\begin{eqnarray}
F_{0}(x) &=& -x + x^2 \arcsin^2(1/\sqrt{x})\,,
\nonumber\\
F_{1/2}(x) &=& -2 x-2x(1-x)\arcsin^2(1/\sqrt{x})\,,
\\
F_1(x) &=& 2 + 3 x + 3 x (2-x) \arcsin^2(1/\sqrt{x})\,.
\nonumber
\label{formfactor}
\end{eqnarray}
The Higgs/stau couplings defined by the Lagrangian interaction term
$v g_{h \tilde{\tau}_i\tilde{\tau}_j}  h \tilde\tau_i ^*\tilde\tau_j/\sqrt{2}$
(where $v=174\GeV$) are explicitly given by 
\bea
g_{h \tilde{\tau}_1\tilde{\tau}_1} &=&
T^\tau_3 \cos^2\theta_{\tilde\tau} - Q_\tau \sin^2\theta_{\rm W}\cos 2\theta_{\tilde\tau} - \frac{m^2_\tau}{M^2_Z}-
\frac{m_\tau \left(A_\ell - \mu \tan\beta\right)}{2 M_Z^2} \sin 2\theta_{\tilde \tau}\,,
\nonumber\\
g_{h \tilde{\tau}_2\tilde{\tau}_2} &=&
T^\tau_3 \sin^2\theta_{\tilde\tau} + Q_\tau \sin^2\theta_{\rm W}\cos 2\theta_{\tilde\tau} - \frac{m^2_\tau}{M^2_Z} +
\frac{m_\tau \left(A_\ell - \mu \tan\beta\right)}{2 M_Z^2} \sin 2\theta_{\tilde \tau}\,,
\label{eq_hstaustau}
\eea
where the stau masses ($m_{{\tilde\tau}_{i}}$) and mixings ($\theta_{\tilde \tau}$) can
be expressed in terms of the left and right soft masses ($m_{L,R}$), which we assume to
be universal for the three generations of sleptons, as
\be
\cos 2\theta_{\tilde \tau} =
\frac{m^2_L - m^2_R}{m^{2}_{{\tilde\tau}_{1}}-m^{2}_{{\tilde\tau}_{2}}}\,,\qquad
\sin 2\theta_{\tilde \tau} =
\frac{2m_\tau\left(A_\ell - \mu\tan\beta\right)}{m^{2}_{{\tilde\tau}_{1}}-m^{2}_{{\tilde\tau}_{2}}}\,,
\label{eq:sin_cos_2theta}
\ee
and
\be
m^{2}_{{\tilde\tau}_{1,2}} =\frac 12\left[
m^2_L + m^2_R \mp \sqrt{(m^2_L - m^2_R)^2 + 4m_\tau^2 (A_\ell-\mu\tan\beta)^2}\right]\,.
\label{mixlep}
\ee
From \eq{gamrat}, for 
$m_{{\tilde\tau}_{2}}\gg m_{{\tilde\tau}_{1}}>m_h/2$ and large $\tan\beta$,
we obtain a  simple expression for the modification of the Higgs decay width into two photons:
\begin{equation}\label{eq:approx}
\frac{\Gamma(h\rightarrow \gamma\gamma)}{\Gamma(h\rightarrow \gamma\gamma)_{\rm SM}}
\approx
\left( 1 + 0.025~\frac{|m_\tau \mu \tan\beta \sin 2\theta_{\tilde \tau}|}{m^{2}_{{\tilde\tau}_{1}}}\right)^2\,.
\end{equation}
Equation~(\ref{eq:approx}) shows that the light stau always increase $\Gamma(h\rightarrow \gamma\gamma)$ and a significant enhancement requires very large values of $\mu\tan\beta$, a stau ${\tilde\tau}_{1}$ as light as possible, and a maximum value for the stau mixing angle ($\sin 2\theta_{\tilde \tau} \approx 1$).
These requirements select a very special region in parameter space, as illustrated in fig.~\ref{fig:gh}a,
where we perform a random scan over
\beq 0< m_L,m_R<\TeV,\qquad -3\TeV< A_\ell, \mu<3\TeV,  \qquad \label{scan}
3<\tan\beta<50 .\eeq
The calculation is performed by keeping the full contribution of staus, without making simplifying approximations such as \eq{eq_hstaustau} or (\ref{eq:approx}). The result
is plotted in fig.~\ref{fig:gh}a as a function of $\Gamma(h\rightarrow \gamma\gamma)$
and $m_{\tilde{\tau}_1}$, for different values of $\mu\tan\beta$.  
The experimental bound on the stau mass from LEP is $m_{\tilde{\tau}_1}> 82-90$~GeV~\cite{stauexp}; lighter staus are allowed only if the difference between the stau and neutralino masses is smaller than a few GeV. The bound on the stau mass implies that the region where $\Gamma(h\rightarrow \gamma\gamma)$ is about twice its SM value corresponds to
$\mu \, (\tan\beta /50) \simgt 2\TeV$.

A Higgs-stau-stau trilinear coupling enhanced by such a large $\mu\tan\beta$ can  lead to an instability of the physical vacuum. 
 In order to understand the origin of the problem, let us consider the relevant terms in the scalar potential
\beq
V= m_{H_{\rm u}}^2 |H_{\rm u}^0|^2+|\mu H_{\rm u}^0-y_\tau {\tilde \tau}_L {\tilde \tau}_R|^2 + \dots \, ,
\label{eqpot}
\eeq
where $y_\tau$ is the tau Yukawa coupling and ${\tilde \tau}_{L,R}$ are the stau fields. The second term in \eq{eqpot} corresponds to the supersymmetric part $|F_{H_{\rm d}}|^2$, and so it is positive-definite and, by itself,
cannot lead to any instability.
The instability for large $\mu$ comes from the fine-tuning required to achieve EW symmetry breaking. Indeed, for large $\tan\beta$, one generally imposes $m_{H_{\rm u}}^2 = - \mu^2 - M_Z^2/2$, and so the first term in \eq{eqpot} becomes large and negative, triggering a deeper minimum at $\vev{H_{\rm u}}\approx \vev{{\tilde \tau}_{L,R}}\approx \mu/y_\tau$. A tree-level analysis of vacuum meta-stability implies the bound~\cite{Rattazzi:1996fb}
\beq
|\mu\tan\beta|  \simlt  39( \sqrt{m_L} + \sqrt{m_R})^2 - 10\TeV.
\label{eq:meta}
\eeq
This result cannot be fully trusted as radiative corrections are large in the region of parameters that we are considering here. Nevertheless, it implies an
important  constraint on the stau contribution to $\Gamma (h\to \gamma\gamma )$. For instance, in the case of maximally mixed staus ($m_L \approx m_R \approx  \sqrt{m_\tau| \mu\tan\beta|}$), we find that \eq{eq:meta} implies $|\mu\tan\beta|  \simlt  20\TeV$. As shown in fig.~\ref{fig:gh}a, this means that a significant enhancement in $\Gamma (h\to \gamma\gamma )$ is possible only for a very light stau.

The large stau trilinear also leads to a correction to the Higgs
 boson mass
\begin{equation}
\left(\delta m^2_h\right)_{\tilde\tau} = 
\frac{v^2 \sin^4\beta}{24 \pi^2}
\left[
\frac{(g^2 + g^{\prime 2}) y_{\tau}^2 \mu^2}{m_{\tilde\ell}^2}-
\frac{ y_{\tau }^4 \mu^4}{2 m_{\tilde \ell}^4}
\right] \, ,
\label{eq:mh_correction}
\end{equation}
where $v=174~$GeV and we have assumed for simplicity $m_L = m_R = m_{\tilde\ell}$.
In the region of parameter space where large effects to $h\to\gamma\gamma$ are
generated, $\left(\delta m^2_h\right)_{\tilde\tau}$ is negative. However, the two
terms in eq.~(\ref{eq:mh_correction}) tend to cancel and the total contribution
is not necessarily large.

In conclusion, a significant enhancement of $\Gamma(h\rightarrow \gamma\gamma)$ selects
a very special region of supersymmetric parameters, with the following characteristics.
The particle ${\tilde{\tau}_1}$ must correspond to a maximally mixed state with mass below $100$~GeV, possibly evading the LEP bound because of an approximate degeneracy with the lightest neutralino.
Higgsinos are very heavy, with masses exceeding the TeV. The Bino must be lighter than ${\tilde{\tau}_1}$, if we impose the condition of a neutral LSP. The Wino, gluino, and squarks are not constrained by these considerations, but must be sufficiently heavy to avoid LHC bounds and to explain the Higgs mass. Of course, this spectrum is not
``natural" in the technical sense, but here we are just following the lead of
experimental data, rather than relying on theoretical considerations.

\section{The anomalous magnetic moment of the muon}
In this section we explain how the special supersymmetric parameters needed to give
a large enhancement of $\Gamma(h\rightarrow \gamma\gamma)$ lead to a well-defined prediction for
$\delta a_\mu$, under the assumption that the soft terms in the lepton sector are universal.

The leading supersymmetric contributions to $\delta a_\mu$
are captured by the following approximate expression~\cite{Moroi:1995yh}
\bea
\label{eq:gm2_MI}
\delta a_\mu &=&
\frac{\alpha \, m^2_\mu \, \mu\,M_{2} \tan\beta}{4\pi \sin^2\theta_W \, m_{L}^{2}}
\left[ \frac{f_{\chi}(M_{2}^2/m_{L}^2)-f_{\chi}(\mu^2/m_{L}^2)}{M_2^2-\mu^2} \right]
\nonumber\\
&+&
\frac{\alpha \, m^2_\mu \, \mu\,M_{1} \tan\beta}{4\pi \cos^2\theta_W \, (m_{R}^{2} - m_{L}^{2})}
\left[\frac{f_{N}(M^2_1/m_{R}^2)}{m^2_R} - \frac{f_{N}(M^2_1/m_{L}^2)}{m^2_L}\right] \,,
\eea
where the loop functions are
\bea
f_{\chi}(x) &=& \frac{x^2 - 4x + 3 + 2\ln x}{(1-x)^3}~,\qquad ~f_{\chi}(1)=-2/3, \\
f_{N}(x) &=& \frac{ x^2 -1- 2x\ln x}{(1-x)^3}\,,\qquad\qquad f_{N}(1) = -1/3 \, .
\eea
The first contribution in eq.~(\ref{eq:gm2_MI}) comes from chargino exchange with
an underlying Higgsino/Wino mixing and it decouples for large $\mu$.
Instead, the second term of eq.~(\ref{eq:gm2_MI}) arises from pure Bino exchange
with an underlying smuon left-right mixing and therefore it grows with $\mu$.

Since large corrections to the rate of $h\to\gamma\gamma$ can arise only in the
presence of large left-right mixing terms, hereafter we will focus on the limit
$\mu\gg m_{L,R},M_{1,2}$. In such a limit, assuming for illustrative purposes a
common slepton/gaugino soft mass ${\tilde m} = m_{L,R} = M_{1,2}$, we find
\be
\delta a_\mu \approx 2.8\times 10^{-9}
\frac{\tan\beta}{20}
\left(\frac{300~{\rm GeV}}{\tilde m}\right)^2
\left[\frac{1}{8}
\frac{10}{\mu/{\tilde m}} +
 \frac{\mu/{\tilde m}}{10}
\right]
\,.
\label{eq:gm2_susy_1}
\ee
This shows that the second term of eq.~(\ref{eq:gm2_MI}) provides indeed the dominant effects in the region of the parameter space we are interested in. This is even more
true when the Wino and/or the sneutrino masses suppress the chargino-mediated loop contribution.

We perform the same scan over supersymmetric parameters as before, where now also the gaugino masses $M_{1,2}$ vary up to 3~TeV. We also require a neutral LSP and impose
the experimental limits on charginos. Using exact expressions in the mass eigenstate basis~\cite{Moroi:1995yh}, we show in fig.~\ref{fig:gh}b the correlation between $\Gamma(h\rightarrow \gamma\gamma)/\Gamma(h\rightarrow \gamma\gamma)_{\rm SM}$ and $\delta a_\mu$.
Whenever the diphoton Higgs decay rate deviates significantly from the
SM expectation (by about 40\% or more), $|\delta a_\mu|$ is determined rather
sharply and the prediction coincides with the measured anomaly, provided that
the $\mu$-term has the appropriate sign.

The reason for the sharp correlation lies in the fact that the slepton parameters are almost completely determined by the condition of maximizing their contribution to $\Gamma(h\rightarrow \gamma\gamma)$. 
Once we accept lepton universality and a neutral LSP, the contribution to $\delta a_\mu$ is also essentially fixed. It is an interesting coincidence that the predicted
value of $\delta a_\mu$ agrees with the observed discrepancy.

We have verified that the parameters corresponding to the points plotted in fig.~\ref{fig:gh}b satisfy the bounds from electroweak data. In particular the stau contributions to $\Delta \rho$ is smaller than $2\times 10^{-3}$,
and explicitly  given by
\beq
\Delta \rho = \frac{G_{\rm F}}{4\sqrt{2}\pi^2} \left[ \sin^2\theta_{\tilde \tau} f(m_{\tilde \nu}^2, m_{{\tilde \tau}_1}^2)+ \cos^2\theta_{\tilde \tau} f(m_{\tilde \nu}^2, m_{{\tilde \tau}_2}^2)-\sin^2\theta_{\tilde \tau}\cos^2\theta_{\tilde \tau} f(m_{{\tilde \tau}_1}^2, m_{{\tilde \tau}_2}^2) \right] \,,
\eeq
where
\beq
f(x,y)=\frac{x+y}{2}+\frac{xy}{x-y}\ln \frac{y}{x}\, .
\eeq

Another interesting consequence of the special supersymmetric parameters singled out
by a large enhancement of $\Gamma(h\rightarrow \gamma\gamma)$ is the dark matter relic abundance. The requests that ${\tilde \tau}_1$ is as light as allowed by experimental constraints and that the Bino is
the LSP squeezes the allowed mass range of the two particles, making them near degenerate. Under this condition, coannihilation processes cannot be neglected.

In fig.~\ref{fig:gh}b, green (red) does correspond to a $\tilde\tau$ heavier
than $100$ $(80)\GeV$, and black dots correspond to a lighter $\tilde\tau$, which
is experimentally allowed only if it is quasi-degenerate to a neutralino.
In the region where $\Gamma(h\rightarrow \gamma\gamma)$ is enhanced, the LSP thermal
relic density is typically consistent with dark matter observations. The requirement
of a correct dark-matter density does not further sharpen the prediction
of $\delta a_\mu$.

Possible effects in $\delta a_\mu$ in presence of large deviations in $\Gamma(h\rightarrow \gamma\gamma)$ have already been pointed out in ref.~\cite{hggSUSY}. However, the authors of~\cite{hggSUSY} focused on the chargino effect which, as shown here, is generally subdominant to the neutralino effect in the region where $\Gamma(h\rightarrow \gamma\gamma)$ is strongly enhanced. 
Our point here is not only that a large enhancement of the diphoton Higgs decay rate is compatible with a deviation in the
muon $g-2$, but rather that it almost necessarily implies a value of $\delta a_\mu$
within the experimentally preferred region.

The correlation between large effects in $\Gamma(h\rightarrow \gamma\gamma)$ and
$\delta a_\mu$ is fairly robust, but it relies on several hypotheses that we state
here and comment upon.

\begin{enumerate} 

\item
The soft terms must be (at least approximately) universal in the slepton sector, so that we can relate the stau parameters (entering the diphoton rate) with the smuon parameters (entering $\delta a_\mu$).
This assumption is reasonable, given the strong constraints from lepton-flavor violating processes, such as $\mu \to e \gamma$ and $\tau \to \mu (e) \gamma$. In the limit of large $\tan \beta$, the tau Yukawa coupling is an important source of flavor non-universality. We have checked that renormalization-group effects do not modify our conclusions, even if lepton-flavor universality is assumed at the GUT scale, rather than at the weak scale. However, we also remark that universality could be badly violated if slepton soft masses are diagonal, but not proportional to the identity, in the basis of diagonal lepton Yukawa matrix. This alignment is possible in certain models with global flavor symmetry.

\item
The LSP must be neutral, so that the Bino mass is forced to be lighter than the stau. If this hypothesis did not hold, then the effect in $\delta a_\mu$ could decouple (in the limit $M_1\to \infty$) even in presence of large corrections to $\Gamma(h\rightarrow \gamma\gamma)$. The hypothesis is especially justified in view of dark matter. Indeed, we have explained above how the combination of light Bino and stau allows to account for the correct thermal relic abundance of the LSP. We should also remark that the correlation determines $|\delta a_\mu |$, but not its sign. However, a large deficit in $\delta a_\mu$ is ruled out by data.

\item
The slepton soft parameters $m_{L,R}$ must not exceed the TeV. This hypothesis is important for the correlation because one could consider the limit $m_{L,R}, \mu\to \infty$, while keeping $m_{{\tilde\tau}_{1}}$ fixed. In this limit, $\delta a_\mu$ decouples even if $\Gamma(h\rightarrow \gamma\gamma)$ receives large corrections. 
However, this situation is excluded by the meta-stability bound,
see eq.~(\ref{eq:meta}), which implies that a large effect in $h\to\gamma\gamma$ is obtained for
 $m_{L,R} \lesssim 300\GeV$.
\end{enumerate}

\section{Other Higgs decay modes}
In this section we discuss how the light ${\tilde \tau}_1$ affects
other decay modes of the Higgs.

\subsection*{$h\rightarrow Z\gamma$}
In general, one expects that any state that contributes at the quantum level
to $\Gamma(h\to \gamma\gamma)$ gives a similarly important correction to
$\Gamma(h\to Z\gamma)$, which is given by~\cite{djouadi,Zgamma}
\be
\label{eq:Zgageneral}
\Gamma(h\to Z\gamma)=
\frac{\alpha^3\, m^3_h}{128\pi^2\sin^2\theta_{\rm W}M^2_W}
\left(1-\frac{M_Z^2}{m_h^2}\right)^3
\left| A_{\rm SM} + \sum_{i,j=1}^2
g_{h \tilde{\tau}_i\tilde{\tau}_j} g_{Z \tilde{\tau}_i\tilde{\tau}_j}
Q_{\tau}
\frac{M^2_Z}{m_{\tilde{\tau}_i}m_{\tilde{\tau}_j}}
F_{0}(m_{\tilde{\tau}_i}, m_{\tilde{\tau}_j}) \right|^2 \ ,
\ee
where
\be
A_{\rm SM} = \cot\theta_{\rm W}\, A_1(\frac{4M_W^2}{m_h^2},\frac{4M_W^2}{M_Z^2}) + \frac{ N_c Q_t (\frac{1}{2}T_3^{(t)}-
Q_t \sin^2\theta_{\rm W})}
{\sin\theta_{\rm W}\cos\theta_{\rm W}} F_{1/2}(\frac{4m_t^2}{m_h^2},\frac{4m_t^2}{M_Z^2}) \,. 
\ee
The stau couplings to the $Z$ boson, including mixing between left--
and right--handed  sfermions, are given by
\bea
g_{Z \tilde{\tau}_1 \tilde{\tau}_1}  & = &
\frac{1}{\sin\theta_{\rm W}\cos\theta_{\rm W}}
\left( T^\tau_3 \cos^2\theta_{\tilde\tau} - Q_\tau \sin^2\theta_{\rm W} \right)\,,
\nonumber\\
g_{Z \tilde{\tau}_2 \tilde{\tau}_2}  & = &
\frac{1}{\sin\theta_{\rm W}\cos\theta_{\rm W}}
\left( T^\tau_3 \sin^2\theta_{\tilde\tau} - Q_\tau \sin^2\theta_{\rm W} \right)\,, \\
g_{Z \tilde{\tau}_1 \tilde{\tau}_2}  & = & -T_3^\tau \frac{\sin\theta_{\tilde\tau}\cos\theta_{\tilde\tau}}{\sin\theta_{\rm W}\cos\theta_{\rm W}}\,,
\nonumber
\label{eq:Zstaustau}
\eea
where $Q_\tau=-1$, $T_3^\tau = -1/2$, $Q_t=2/3$, $N_c=3$ and the SM loop functions are
\bea
\label{eq:loop1}
A_1(x,y)&=& 4 (3-\tan^2\theta_{\rm W}) I_2(x,y)+ \left[ (1+2x^{-1}) \tan^2\theta_{\rm W} - (5+2x^{-1})\right] I_1(x,y)\,,  \\
\label{eq:loop5}
 F_{1/2}(x,y)& =& 4[I_1(x,y)-I_2(x,y)]\,,
\eea
\bea
I_1(x,y) &=& \frac{x y}{2(x-y)} + \frac{x^2 y^2}{2(x-y)^2}[ f(x)-f(y)] +
\frac{x^2 y}{(x-y)^2}[g(x)-g(y)]\,, \\
I_2(x,y) &=& - \frac{x y}{2(x-y)} [ f(x)-f(y)]\,, \
\eea
\be
 f(x) = \arcsin^2 \sqrt{1/x} \ , \qquad
 g(x) = \sqrt{x -1} \arcsin \sqrt{1/x} \ .
\ee
The scalar function $F_0(m_1,m_2) $ is written in terms of Passarino-Veltman $C$-functions~\cite{djouadi}, with the simple limiting case
$F_{0}(m,m) = 2 I_1(4{m^2}/{m_h^2}, 4{m^2}/{M_Z^2})$. 
In the limit of massless $Z$ boson, all the loop functions for $h\to Z \gamma$
reduce to the corresponding ones for $h\to \gamma\gamma$.

\begin{figure}[tb]
\begin{center}
$$\includegraphics[width=0.45\textwidth]{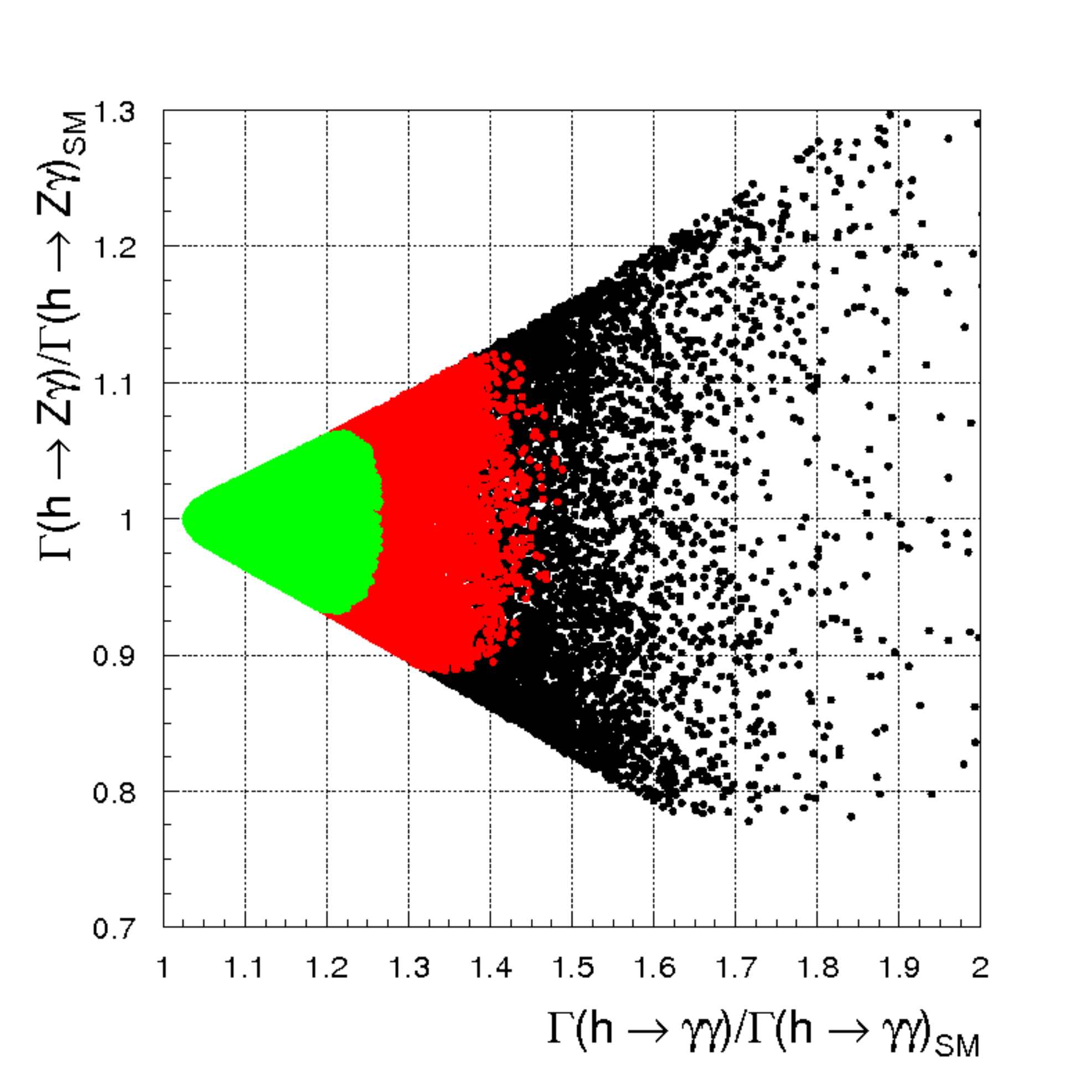}$$
\caption{\em Correlation between $\Gamma(h\to\gamma\gamma)$ and $\Gamma(h\to Z\gamma)$. The scan over the supersymmetric parameters is as in fig.~\ref{fig:gh}
Green (red) does correspond to a $\tilde\tau$ heavier than $100$ $(80)\GeV$,
and black dots correspond to a lighter $\tilde\tau$, which is experimentally
allowed  if it is quasi-degenerate to a neutralino.
\label{fig:hdecays}}
\end{center}
\end{figure}

In our case $h\to Z\gamma$ is generated at one-loop 
level by three diagrams with $\tilde\tau_1\tilde\tau_1$, $\tilde\tau_2\tilde\tau_2$
and $\tilde\tau_1\tilde\tau_2$ in the loop. The first two diagrams are  suppressed
because in the limit of maximal stau mixing the $Z\tilde{\tau}_i\tilde{\tau}_i$
couplings, see \eq{eq:Zstaustau}, are proportional to $1-4\sin^2\theta_{\rm W}$,
which is accidentally small. The latter $\tilde\tau_1\tilde\tau_2$ diagram is also suppressed by the mass of the $\tilde\tau_2$ state and by the $h \tilde\tau_1 \tilde\tau_2$
coupling, which vanishes in the limit of maximal stau mixing (see also~\cite{Zgamma}).
Thereby, the correction to $\Gamma(h\to Z\gamma)$ is smaller than the one to $\Gamma(h\to \gamma \gamma)$. Moreover, the leading effect (coming from ${\tilde \tau}_1$ exchange)
can be either positive or negative, as it is proportional to $g_{Z \tilde{\tau}_1 \tilde{\tau}_1}\propto \sin^2\theta_W -\cos^2\theta_{\tilde \tau}/2$: it is negative
if right-handed sleptons are lighter than left-handed ones ($m_R< m_L$) and positive otherwise. The result of our scan over supersymmetric parameters showing the
correlation between  $\Gamma(h\to Z\gamma)$ and  $\Gamma(h\to \gamma \gamma)$
is presented in \fig{fig:hdecays}.

\subsection*{$h\to\tau\tau$}
In the limit of large $\tan\beta$, there are sizable one-loop corrections to the
$\tau$ mass~\cite{RS}. Including only  the leading effect of the light stau and 
Bino we obtain
\beq \Delta m_\tau =  \frac{\alpha \sin2\theta_{\tilde\tau} }{8\pi \cos^2\theta_{\rm W}}
M_{1}[ B(m_{\tilde\tau_2}^2,M_1^2)-B(m_{\tilde\tau_1}^2,M_1^2) ]
\,,
\label{corrtau}
\eeq
where $B(x,y)=(x\ln x- y \ln y)/(x-y)$. So
\beq \frac{\Delta m_\tau}{m_\tau} \approx 20\% ~  \frac{\mu}{\TeV}~\frac{\tan\beta}{50}~ \frac{100\GeV}{m_{{\tilde \tau}_1}}.
\eeq
The same one-loop diagrams that modify the $\tau$ mass according to \eq{corrtau}
also affect the process $h\to \tau\tau$. After expressing the result in terms
of the physical tau mass, the effect is given by the difference of these diagrams
when evaluated on-shell with respect to when evaluated  at zero momentum.
Including only the effect of the lightest stau we find, expanding at first order
in $m_h^2/4m_{\tilde{\tau}_1}^2$,

\beq 
\frac{\Gamma(h\to\tau\tau)}{\Gamma(h\to\tau\tau)_{\rm SM}}= \left[1 +
 \frac{\alpha\, \mu\tan\beta}{384\pi \cos^2\theta_{\rm W}} 
\sin^22\theta_{\tilde\tau}\,  M_1\,   \frac{m_h^2}{m_{\tilde{\tau}_1}^4}
f(\frac{M_1^2}{m^2_{\tilde{\tau}_1}})
 \right]^2\,,
\eeq
where $f(x)=2(1-6x+3x^2+2x^3-6x^2\ln x)/(1-x)^4$ such that $f(1)=1$.
The correction to $h\to\tau\tau$ is very small, at the level of a few $\%$.

\begin{figure}[t]
$$
\includegraphics[width=0.5\textwidth]{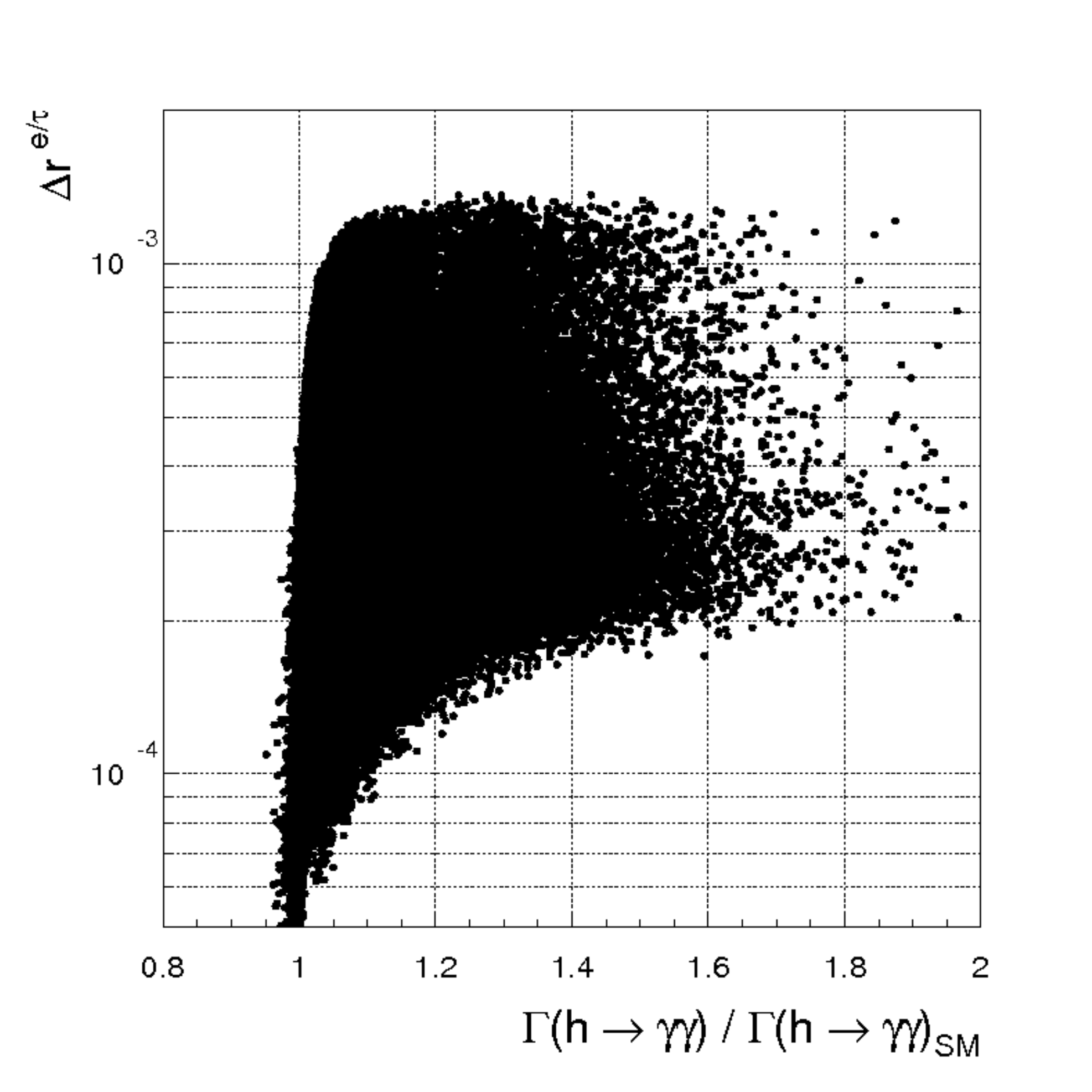}\qquad \includegraphics[width=0.5\textwidth]{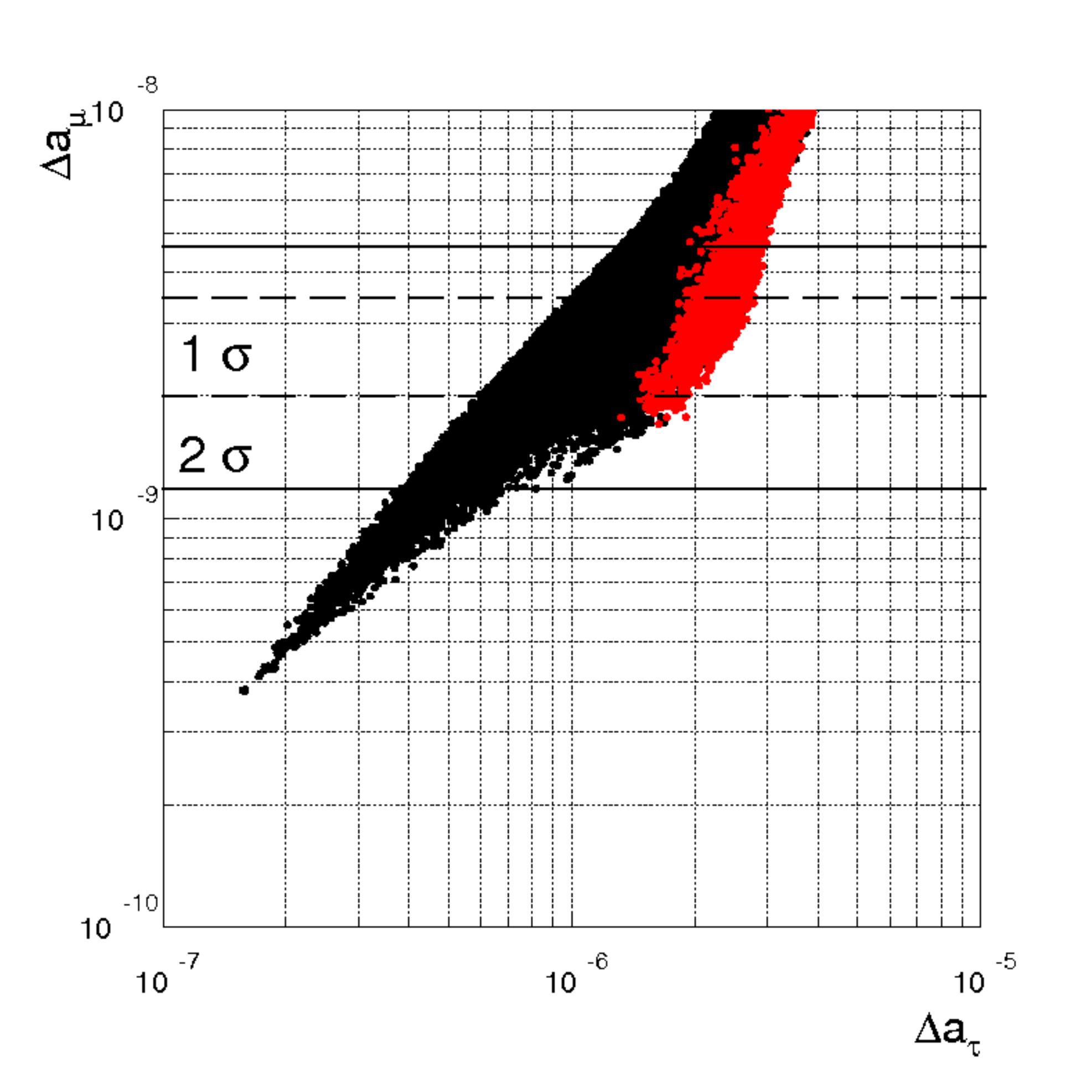}$$
\caption{\em {\bf Left:}
Lepton-flavor universality breaking effects in $\tau$ decays described
by the quantity $\Delta r^{\ell /\tau}$, see eq.~(\ref{eq:delta_r}),
as a function of $\Gamma(h\to\gamma\gamma)$ normalized to its SM value.
{\bf Right}: $\delta a_\mu$ vs. $\delta a_\tau$. Red points correspond to
the currently favored region for $\Gamma(h\rightarrow \gamma\gamma)$.
\label{fig:lfu}}
\end{figure}

\section{Violation of lepton flavor universality}

The most important correction to $\Gamma(h\to\gamma\gamma)$ in supersymmetry comes
from the left-right mixing in the slepton mass matrix. Such a contribution breaks
the EW gauge symmetry and is lepton flavor non-universal since it is proportional
to the $\tau$ Yukawa. Therefore, it is natural to expect violations of lepton flavor
universality in various high-energy and low-energy processes. In this section,
we study the implications of an enhancement of $\Gamma(h\to\gamma\gamma)$ for such
processes.

Lepton universality has been probed at the few per-mill level so far, in processes
such as $P\to\ell\nu$, $\tau\to P\nu$ (where $P=\pi,K$), $\ell_i\to \ell_j\bar\nu\nu$, $Z\to\ell\ell$ and $W\to\ell\nu$. Since the sources of lepton non-universality relevant
for our discussion are proportional to the lepton Yukawa couplings, only $\mu/\tau$ and
$e/\tau$ but not $e/\mu$ universalities will be affected. The relevant processes probing
the $\mu/\tau$ and $e/\tau$ sectors and their experimental situation is summarized in table~\ref{table:LFU}.

Taking for example the process $\tau\to\ell\nu\bar\nu$, we can define the quantity
\beq
\frac{(R^{\ell/\tau})}{(R^{\ell/\tau})_{\rm SM}}
= 1 + \Delta r^{\ell/\tau}\,,\qquad \ell = e,\mu~.
\label{eq:delta_r}
\eeq
Here $(R^{\mu/\tau})_{\rm SM}=\Gamma(\tau\to e\nu\bar\nu)_{\rm SM}/\Gamma(\mu\to e\nu\bar\nu)_{\rm SM}$ and $(R^{\mu/\tau})=\Gamma(\tau\to e\nu\bar\nu)/\Gamma(\mu\to e\nu\bar\nu)$ so that
$\Delta r^{\mu/\tau}\neq 0$ signals the presence of new physics violating lepton universality. At tree level, gauge invariance guarantees lepton flavor universality
of the $W$ interactions. This universality is maintained to all orders for exact
$SU(2)$ gauge symmetry, while it is broken in general at the loop level after EW
breaking. In the effective field theory language, this corresponds to effects
induced, after EW symmetry breaking, by gauge invariant dimension-six operators
such as $(\bar{L}_L \gamma^\mu L_L)(H^\dagger D_\mu H)$.

In order to describe the above effects, it is convenient to consider
the following effective Lagrangian
\begin{eqnarray}
{\cal L}_{\rm eff} =
\overline{\ell}_{L} Z_{L}^{\ell} i\!\not\!\partial~\ell_{L} +
\overline{\nu}_{L} Z^{\nu}_{L} i\!\not\!\partial~\nu_{L}-\frac{g}{\sqrt2} W_\mu^-
\overline{\ell}_{L}\gamma^\mu Z^{W}_{L}\nu_{L} +{\rm h.c.}~,
\end{eqnarray}
where the $Z^a$ matrices can be written as
\beq
\left(Z^{a}\right)_{ij}= \delta_{ij} + \left(\eta^{a}\right)_{ij}\,,
\qquad a=\nu,\ell,W\,.
\eeq
\begin{table}[t]
\centering
\renewcommand{\arraystretch}{1.5}
\begin{tabular}{lcc}
\hline\hline
Channel & $\Delta r^{\mu /\tau} $ \\
\hline
\hline
$\Gamma(\pi \rightarrow \mu \,\bar{\nu}_\mu)/ \Gamma(\tau  \rightarrow \pi \, \nu_\tau)$ & $0.016\pm 0.008 $~\cite{BABAR-Univ:10} \\
$\Gamma(K \rightarrow \mu \, \bar{\nu}_\mu)/\Gamma(\tau  \rightarrow K \, \nu_\tau)$ &
$0.037 \pm 0.016 $~\cite{BABAR-Univ:10} \\
$\Gamma(Z \rightarrow \mu^+ \mu^-)/\Gamma(Z \rightarrow \tau^+ \tau^-)$ &
$-0.0011 \pm 0.0034$~\cite{PDG,LEPEWWG,LEPEWWG_SLD:06,Abbaneo:2001ix} \\
$\Gamma(W \rightarrow \mu \, \bar{\nu}_\mu)/\Gamma(W \rightarrow \tau \, \bar{\nu}_\tau )$ &
$ -0.060 \pm 0.021 $~\cite{PDG,LEPEWWG,LEPEWWG_SLD:06,Abbaneo:2001ix} \\
 $\Gamma(\mu \rightarrow \nu_\mu\, e\, \bar{\nu}_e ) / \Gamma(\tau \rightarrow \nu_\tau\, e \,\bar{\nu}_e)$ & $ -0.0014 \pm 0.0044 $~\cite{BABAR-Univ:10} \\
\hline\hline
Channel & $\Delta r^{e /\tau} $ \\
\hline
\hline
$\Gamma(Z \rightarrow e^+ e^-)/\Gamma(Z \rightarrow \tau^+ \tau^-)$ &
$-0.0020 \pm 0.0030$~\cite{PDG,LEPEWWG,LEPEWWG_SLD:06,Abbaneo:2001ix} \\
$\Gamma(W \rightarrow e \, \bar{\nu}_e)/\Gamma(W \rightarrow \tau \, \bar{\nu}_\tau )$ &
$ -0.044 \pm 0.021 $~\cite{PDG,LEPEWWG,LEPEWWG_SLD:06,Abbaneo:2001ix} \\
$\Gamma(\mu \rightarrow \nu_\mu\, e\,\bar{\nu}_e ) /
\Gamma(\tau \rightarrow \nu_\tau\,\mu\,\bar{\nu}_\mu)$ & $ -0.0032 \pm 0.0042 $~\cite{BABAR-Univ:10} \\
\hline\hline
\end{tabular}
\renewcommand{\arraystretch}{1}
\caption{Experimental constraints on $\Delta r^{e/\tau}$ and $\Delta r^{\mu/\tau}$.}
\label{table:LFU}
\end{table}
The Hermiticity of the Lagrangian ensures that
$(\eta^{\ell,\nu})^\dagger=\eta^{\ell,\nu}$, while $\eta^W$ is general.
After rescaling the lepton fields to make their kinetic terms canonical
\begin{eqnarray}
\nu_{L}
\rightarrow
\left(1 - {1\over2}\eta_{L}^{\nu}\right)
\nu_{L},
\qquad
\ell_{L}
\rightarrow
\left(1 - {1\over2}\eta_{L}^{\ell}\right)
\ell_{L}\,,
\label{eqn:rescaling_1loop}
\end{eqnarray}
the $W$-boson interaction becomes
\beq
{\cal L}_{int}=
-\frac{g}{\sqrt{2}}W^{-}_\mu
\overline{\ell}_L\gamma^\mu Z_W
\nu_L +{\rm h.c.}\,,
\eeq
\beq
Z_W = 1 + \eta^W_L - \frac{\eta^{\ell}_{L}+\eta^{\nu}_{L}}{2}\,.
\label{eq:Wln}
\eeq
Therefore, we can write $\Delta r^{\ell/\tau}$ in terms of $Z_W$,
\beq
\Delta r^{\ell/\tau} = \frac{\left|Z^{\ell\ell}_W\right|^2}{\left|Z^{\tau\tau}_W\right|^2} -1 \, .
\label{eq:deltar_vs_Wln}
\eeq
The calculation of $\Delta r^{\ell/\tau}$ in neutral-current processes is analogous,
although the expression is more complicated since left and right couplings can receive
different corrections.

In the case of universal soft terms, out of the various sources of EW breaking felt by supersymmetric particles ($D$-terms, gaugino/higgsino mixing terms, and left-right sfermion mixings), only left-right mixings violate lepton universality, as they are proportional
to Yukawa couplings. The value of $\Delta r^{\ell/\tau}$ relevant to charged-current
interactions can be derived from the calculation of $\mu$ decay in supersymmetry~\cite{Chankowski:1993eu,Krawczyk:1987zj} and we have included the full
result in our numerical analysis. The parametric form is
\begin{equation}
\Delta r^{\ell/\tau}
\sim
\frac{\alpha}{4\pi}
\frac{m^2_{\tau}\left|A_{\tau}-\mu\tan\beta\right|^2}
{M^2_{1,2}~m^2_{\tilde{\tau}_1}}~,\qquad \ell = e,\mu~,
\end{equation}
where we picked up a double left-right mixing term for the third slepton generation.
Therefore, we expect effects of lepton non-universality at the per-mill level, whenever 
large contributions to $\Gamma(h\to\gamma\gamma)$ are induced.

This is illustrated in fig.~\ref{fig:lfu}, where we show the correlation
between $\Gamma(h\to\gamma\gamma)$ and $\Delta r^{\ell/\tau}$ making the
same scan as before. As we can see, large effects in $h\to\gamma\gamma$ (40\% or more) would unambiguously imply non-universality effects in $\tau$ decays at the level of
$0.2\times10^{-3}\lesssim\Delta r^{\ell/\tau}\lesssim 2\times10^{-3}$. These values are
well within the expected sensitivity of a SuperB machine~\cite{Bona:2007qt}.
Note that, within our scenario, the predicted effects in the $\mu/\tau$ and $e/\tau$
sectors are the same, as we are modifying only the gauge-boson vertex with $\tau$,
but not with $e$ or $\mu$.

Finally, we comment on the stau contribution to the $\tau$ anomalous magnetic moment,
which is given by
\beq
\delta a_\tau = K_{\tilde \tau}\, \frac{m_\tau^2}{m_\mu^2}\, \delta a_\mu .
\eeq
Here $K_{\tilde \tau}$ is a coefficient equal to one in the case in which stau and smuon
states are degenerate in mass. In the case of a light stau, $K_{\tilde \tau}>1$ and, for
parameters giving a significant enhancement of $\Gamma(h\rightarrow \gamma\gamma)$, we
find that $\delta a_\tau$ can be as large as $3\times 10^{-6}$.
This is shown in fig.~\ref{fig:lfu} where we present $\delta a_\mu$ versus $\delta a_\tau$, with the
red points corresponding to the currently favored region for $\Gamma(h\rightarrow \gamma\gamma)$. Observation of these values of $a_\tau$ is a difficult experimental challenge.


\section{Conclusions}

Now that the Higgs has been discovered,  the attention turns towards its couplings,
which carry important information about the nature of the new particle. The preliminary indication for an excess in the di-photon rate, together with $ZZ^*$ and $WW^*$ rates consistent with the SM, has triggered exploratory studies of new-physics effects~\cite{fit,hggSUSY,hgg}. In this paper we have pointed a strong correlation, which is present in a broad class of supersymmetric theories, between large contributions to $\Gamma(h\to\gamma\gamma)$ and the observed discrepancy in the magnetic moment of the muon.

In the context of supersymmetry, a confirmation of the preliminary results on the Higgs couplings would point towards a rather peculiar (and technically ``unnatural") choice of parameters. The spectrum should contain a light and maximally mixed stau 
with mass below 100~GeV, together with heavy higgsinos (with masses above 1 TeV) and a light Bino as the LSP. Vacuum metastability imposes a powerful constraint on the stau parameters. Thus, a large enhancement in $\Gamma(h\to\gamma\gamma)$ requires extreme values of the supersymmetric parameters, which essentially fix all slepton soft masses. Consequently, the value of $\delta a_\mu$ is nearly determined too, if we assume slepton universality. Remarkably, the prediction for $\delta a_\mu$ turns out in perfect agreement with the measurement of the magnetic moment of the muon, once we rely on the latest SM determinations~\cite{gm2mth}.

The special parameters singled out by a large enhancement of $\Gamma(h\to\gamma\gamma)$ have other consequences, beyond $\delta a_\mu$. {\it (i)} The request of a neutral LSP corners the Bino to have the right properties to account for dark matter, through Bino-stau coannihilation. {\it (ii)} In spite of the similarity between the corresponding Feynman diagrams, $h\to Z\gamma$ is less affected than $h\to \gamma \gamma$, because of an accidental suppression of the $Z$ coupling with staus. {\it (iii)} Lepton flavor universality is broken in the $\tau$ sector at the level of $10^{-3}$, predicting several observable effects in future experiments. 

Discovering new physics through virtual effects is always a difficult task.
The intriguing correlation, pointed out in this paper, between a  large enhancement in $\Gamma(h\to\gamma\gamma)$ and the magnetic moment of the muon can provide an interesting testing ground for evidence of new physics. This correlation is particularly striking in the case of supersymmetry, but can be present in other contexts, for instance in models with new vector-like fermions~\cite{preparation}.


\subsubsection*{Acknowledgments}
We thank Teppei Kitahara, Norimi Yokozaki, Satoshi Shirai for useful comments. This work was supported by the ESF grant MTT8 and by
SF0690030s09 project.

\appendix
 
\end{document}